\documentclass[twocolumn]{aastex631}

\usepackage{xspace}
\usepackage{multirow}

\usepackage[encapsulated]{CJK}

\newcommand*{\JWST}{\emph{JWST}\xspace}
\newcommand*{\HST}{\emph{HST}\xspace}
\newcommand*{\Spitzer}{\emph{Spitzer}\xspace}
\newcommand{\minus}{\scalebox{0.5}[1.0]{$-$}}

\shorttitle{The \JWST UNCOVER Treasury Survey}
\shortauthors{Bezanson et al.}

\begin{document}

\title{The \JWST UNCOVER Treasury survey:\\Ultradeep NIRSpec and NIRCam ObserVations before the Epoch of Reionization}

\author[0000-0001-5063-8254]{Rachel Bezanson}
\affiliation{Department of Physics and Astronomy and PITT PACC, University of Pittsburgh, Pittsburgh, PA 15260, USA}
\author[0000-0002-2057-5376]{Ivo Labbe}
\affiliation{Centre for Astrophysics and Supercomputing, Swinburne University of Technology, Melbourne, VIC 3122, Australia}

\author[0000-0001-7160-3632]{Katherine E. Whitaker}
\affiliation{Department of Astronomy, University of Massachusetts, Amherst, MA 01003, USA}
\affiliation{Cosmic Dawn Center (DAWN), Niels Bohr Institute, University of Copenhagen, Jagtvej 128, K{\o}benhavn N, DK-2200, Denmark} 
\author[0000-0001-6755-1315]{Joel Leja}
\affiliation{Department of Astronomy \& Astrophysics, The Pennsylvania State University, University Park, PA 16802, USA}
\affiliation{Institute for Computational \& Data Sciences, The Pennsylvania State University, University Park, PA 16802, USA}
\affiliation{Institute for Gravitation and the Cosmos, The Pennsylvania State University, University Park, PA 16802, USA}

\author[0000-0002-0108-4176]{Sedona H. Price}
\affiliation{Department of Physics and Astronomy and PITT PACC, University of Pittsburgh, Pittsburgh, PA 15260, USA}
\author[0000-0002-8871-3026]{Marijn Franx}
\affiliation{Leiden Observatory, Leiden University, P.O.Box 9513, NL-2300 AA Leiden, The Netherlands}
\author[0000-0003-2680-005X]{Gabriel Brammer}
\affiliation{Cosmic Dawn Center (DAWN), Niels Bohr Institute, University of Copenhagen, Jagtvej 128, K{\o}benhavn N, DK-2200, Denmark}
\author[0000-0001-9002-3502]{Danilo Marchesini}
\affiliation{Department of Physics and Astronomy, Tufts University, 574 Boston Ave., Medford, MA 02155, USA}
\author[0000-0002-0350-4488]{Adi Zitrin}
\affiliation{Physics Department, Ben-Gurion University of the Negev, P.O. Box 653, Beer-Sheva 8410501, Israel}

\author[0000-0001-9269-5046]{Bingjie Wang (\begin{CJK*}{UTF8}{gbsn}王冰洁\ignorespacesafterend\end{CJK*})}
\affiliation{Department of Astronomy \& Astrophysics, The Pennsylvania State University, University Park, PA 16802, USA}
\affiliation{Institute for Computational \& Data Sciences, The Pennsylvania State University, University Park, PA 16802, USA}
\affiliation{Institute for Gravitation and the Cosmos, The Pennsylvania State University, University Park, PA 16802, USA}
\author[0000-0003-1614-196X]{John R. Weaver}
\affiliation{Department of Astronomy, University of Massachusetts, Amherst, MA 01003, USA}
\author[0000-0001-6278-032X]{Lukas J. Furtak}
\affiliation{Physics Department, Ben-Gurion University of the Negev, P.O. Box 653, Beer-Sheva 8410501, Israel}

\author[0000-0002-7570-0824]{Hakim Atek}
\affiliation{Institut d'Astrophysique de Paris, CNRS, Sorbonne Universit\'e, 98bis Boulevard Arago, 75014, Paris, France}
\author[0000-0001-7410-7669]{Dan Coe}
\affiliation{Space Telescope Science Institute (STScI), 3700 San Martin Drive, Baltimore, MD 21218, USA}
\affiliation{Association of Universities for Research in Astronomy (AURA), Inc. for the European Space Agency (ESA)}
\affiliation{Center for Astrophysical Sciences, Department of Physics and Astronomy, The Johns Hopkins University, 3400 N Charles St. Baltimore, MD 21218, USA}
\author[0000-0002-7031-2865]{Sam E. Cutler}
\affiliation{Department of Astronomy, University of Massachusetts, Amherst, MA 01003, USA}
\author[0000-0001-8460-1564]{Pratika Dayal}
\affiliation{Kapteyn Astronomical Institute, University of Groningen, P.O. Box 800, 9700 AV Groningen, The Netherlands}
\author[0000-0002-8282-9888]{Pieter van Dokkum}
\affiliation{Department of Astronomy, Yale University, New Haven, CT 06511, USA}
\author[0000-0002-1109-1919]{Robert Feldmann}
\affiliation{Institute for Computational Science, University of Zurich, Winterhurerstrasse 190, CH-8006 Zurich, Switzerland}
\author[0000-0003-4264-3381]{Natascha M. Förster Schreiber}
\affiliation{Max-Planck-Institut f{\"u}r extraterrestrische Physik, Gie{\ss}enbachstra{\ss}e 1, 85748 Garching, Germany}

\author[0000-0001-7201-5066]{Seiji Fujimoto}\altaffiliation{Hubble Fellow}
\affiliation{
Department of Astronomy, The University of Texas at Austin, Austin, TX 78712, USA
}

\author[0000-0002-7007-9725]{Marla Geha}
\affiliation{Department of Astronomy, Yale University, New Haven, CT 06511, USA}
\author[0000-0002-3254-9044]{Karl Glazebrook}
\affiliation{Centre for Astrophysics and Supercomputing, Swinburne University of Technology, Melbourne, VIC 3122, Australia}
\author[0000-0002-2380-9801]{Anna de Graaff}
\affiliation{Max-Planck-Institut f{\"u}r Astronomie, K{\"o}nigstuhl 17, D-69117, Heidelberg, Germany}
\author{Jenny E. Greene}
\affiliation{Department of Astrophysical Sciences, 4 Ivy Lane, Princeton University, Princeton, NJ 08544, USA}
\author[0000-0002-0000-2394]{St\'ephanie Juneau}
\affiliation{NSF’s National Optical-Infrared Astronomy Research Laboratory, 950 N. Cherry Avenue, Tucson, AZ 85719, USA}
\author[0000-0002-3838-8093]{Susan Kassin}
\affiliation{Space Telescope Science Institute (STScI), 3700 San Martin Drive, Baltimore, MD 21218, USA}
\author[0000-0002-7613-9872]{Mariska Kriek}
\affiliation{Leiden Observatory, Leiden University, P.O.Box 9513, NL-2300 AA Leiden, The Netherlands}
\author[0000-0002-3475-7648]{Gourav Khullar}
\affiliation{Department of Physics and Astronomy and PITT PACC, University of Pittsburgh, Pittsburgh, PA 15260, USA}
\author[0000-0003-0695-4414]{Michael Maseda}
\affiliation{Department of Astronomy, University of Wisconsin-Madison, 475 N. Charter St., Madison, WI 53706 USA}
\author[0000-0002-8530-9765]{Lamiya A. Mowla}
\affiliation{Dunlap Institute for Astronomy and Astrophysics, 50 St. George Street, Toronto, Ontario, M5S 3H4, Canada}
\author[0000-0002-9330-9108]{Adam Muzzin}
\affiliation{Department of Physics and Astronomy, York University, 4700 Keele Street, Toronto, Ontario, ON MJ3 1P3, Canada}
\author[0000-0003-2804-0648]{Themiya Nanayakkara}
\affiliation{Centre for Astrophysics and Supercomputing, Swinburne University of Technology, Melbourne, VIC 3122, Australia}
\author[0000-0002-7524-374X]{Erica J. Nelson}
\affiliation{Department for Astrophysical and Planetary Science, University of Colorado, Boulder, CO 80309, USA}
\author[0000-0001-5851-6649]{Pascal A. Oesch}
\affiliation{Department of Astronomy, University of Geneva, Chemin Pegasi 51, 1290 Versoix, Switzerland}
\affiliation{Cosmic Dawn Center (DAWN), Niels Bohr Institute, University of Copenhagen, Jagtvej 128, K{\o}benhavn N, DK-2200, Denmark}
\author[0000-0003-4196-0617]{Camilla Pacifici}
\affiliation{Space Telescope Science Institute (STScI), 3700 San Martin Drive, Baltimore, MD 21218, USA}
\author[0000-0002-9651-5716]{Richard Pan}
\affiliation{Department of Physics and Astronomy, Tufts University, 574 Boston Ave., Medford, MA 02155, USA}
\author[0000-0001-7503-8482]{Casey Papovich}
\affiliation{Department of Physics and Astronomy, Texas A\&M University, College Station, TX, 77843-4242 USA}
\affiliation{George P. and Cynthia Woods Mitchell Institute for Fundamental Physics and Astronomy, Texas A\&M University, College Station, TX, 77843-4242 USA}
\author[0000-0003-4075-7393]{David J. Setton}
\affiliation{Department of Physics and Astronomy and PITT PACC, University of Pittsburgh, Pittsburgh, PA 15260, USA}
\author[0000-0003-3509-4855]{Alice E. Shapley}
\affiliation{Physics \& Astronomy Department, University of California: Los Angeles, 430 Portola Plaza, Los Angeles, CA 90095, USA}
\author[0000-0001-8034-7802]{Renske Smit}
\affiliation{Astrophysics Research Institute, Liverpool John Moores University, 146 Brownlow Hill, Liverpool L3 5RF, UK}
\author[0000-0001-7768-5309]{Mauro Stefanon}
\affiliation{Departament d'Astronomia i Astrofisica, Universitat de Valencia, C. Dr. Moliner 50, E-46100 Burjassot, Valencia, Spain}
\affiliation{Unidad Asociada CSIC ``Grupo de Astrofisica Extragalactica y Cosmologi'' (Instituto de Fisica de Cantabria - Universitat de Valencia)}
\author[0000-0002-5522-9107]{Edward N.\ Taylor}
\affiliation{Centre for Astrophysics and Supercomputing, Swinburne University of Technology, Melbourne, VIC 3122, Australia}
\author[0000-0003-2919-7495]{Christina C. Williams}
\affiliation{NSF’s National Optical-Infrared Astronomy Research Laboratory, 950 N. Cherry Avenue, T            ucson, AZ 85719, USA}
\affiliation{Steward Observatory, University of Arizona, 933 North Cherry Avenue, Tucson, AZ 85721, USA}

\begin{abstract}
In this paper we describe the survey design for the Ultradeep NIRSpec and NIRCam ObserVations before the Epoch of Reionization (UNCOVER) Cycle 1 \JWST Treasury program, which executed its early imaging component in November 2022. The UNCOVER survey includes ultradeep ($\sim29-30\mathrm{AB}$) imaging of $\sim$45 arcmin$^2$ on and around the well-studied Abell 2744 galaxy cluster at $z=0.308$ and will follow-up ${\sim}500$ galaxies with extremely deep low-resolution spectroscopy with the NIRSpec/PRISM during the summer of 2023 {, with repeat visits in summer 2024}. We describe the science goals, survey design, target selection, and planned data releases. We also present and characterize the depths of the first NIRCam imaging mosaic, highlighting previously unparalleled resolved and ultradeep 2-4 micron imaging of known objects in the field. The UNCOVER primary NIRCam mosaic spans 28.8 arcmin$^2$ in seven filters (F115W, F150W, F200W, F277W, F356W, F410M, F444W) and 16.8 arcmin$^2$ in our NIRISS parallel (F115W, F150W, F200W, F356W, and F444W).  To maximize early community use of the Treasury data set, we publicly release full reduced mosaics of public \JWST imaging including 45 arcmin$^2$ NIRCam and 17 arcmin$^2$ NIRISS mosaics on and around the Abell 2744 cluster, including the Hubble Frontier Field primary and parallel footprints.
\end{abstract}


\section{Introduction} \label{sec:intro}

A lasting legacy of the {\em Hubble Space Telescope} (\HST) and \Spitzer \textit{Space Telescope} is the discovery and characterization of galaxies to $z{\sim}11$, looking back 97\% of the time to the Big Bang \citep[e.g.,][]{coe:13,oesch:16}. Extragalactic deep fields with Hubble imaging and grism spectroscopy to $\sim1.6$ microns have provided a reasonably complete census of the cosmic star formation history \citep[e.g.,][]{madau:14, finkelstein:15, Oesch:18,2022ApJ...940...55B}, and combined with \Spitzer imaging, constraints on the mass functions of galaxies since $z=3$ \citep[e.g.,][]{leja:2020,duncan:14,grazian:15,song:16,bhatawdekar:19,kikuchihara:20,furtak:21,stefanon:21}. However, despite 10,000 \HST orbits and tens of thousands of hours of \Spitzer observations, our picture of the Universe remains highly incomplete due to reliance on \HST{}-selection (rest-frame $UV$ at $z{>}4$). This means we were biased towards the youngest, least dusty galaxies, and had been unable to study the first galaxies at $z\gtrsim12$. Understanding these first galaxies is crucial since they mark the start of the process of cosmic reionization and are the first sources of metals in the Universe \citep[e.g.,][]{Dayal:18}. 

The \JWST was specifically designed to address these limitations by providing high-throughput, sensitive imaging and spectroscopy at $1-5\micron$ with the NIRCam  {\citep{2005SPIE.5904....1R,2003SPIE.4850..478R,2023PASP..135b8001R}, NIRISS \citep{2012SPIE.8442E..2RD} and NIRSpec instruments \citep{2022A&A...661A..80J, Rigby:22, 2023PASP..135c8001B}}. These capabilities have begun to enable us to find galaxies deep into the Dark Ages ($z{=}10{-}20$) \citep[e.g.,][]{naidu:22,atek:22,finkelstein:22,donnan:22,harikane:22,2023MNRAS.518.4755A}, to directly study the faint galaxies responsible for reionizing the universe a few hundred million years later, and identifying dusty and quiescent galaxies to low masses to redshift $z\sim10$ \citep[e.g.,][]{carnall:22,labbe:22,naidu:22b,zavala:22,nelson+22}. The intrinsically faintest galaxies can be reached with the aid of gravitational lensing. Observations with \HST along gravitational lensing galaxy clusters \cite[e.g., in the Hubble Frontier Fields,][]{lotz:17} have identified some exceptionally magnified individual objects \citep[e.g.,][]{zitrin:14,kelly:15}, increasing effective depths of e.g., reionization era UV luminosity functions by several magnitudes \citep[e.g.,][]{Atek:18,Ishigaki:18,bouwens:21, Kauffmann:22,2022ApJ...940...55B,2022ApJ...931...81B}. However, at the faintest ends, these studies remain besieged by systematics. 

It is in this context that the Ultradeep NIRSpec and NIRCam ObserVations before the Epoch of Reionization (UNCOVER) Treasury survey was designed (PIs Labb\'e \& Bezanson, \JWST{}-GO\#2561). UNCOVER targets the powerful lensing cluster Abell 2744 and consists of two coordinated components executed in the same cycle. First, a deep NIRCam pre-imaging mosaic in 7 filters for $\sim4-6$ hour per band, and second ultra-deep $3-20$ hour NIRSpec/PRISM low-resolution follow up of NIRCam-detected high-redshift galaxies - each with NIRISS or NIRCam imaging parallels. The imaging and spectroscopy leverages the gravitational lensing boost to push beyond depths achievable in blank fields.  Perhaps the most exciting legacy is the potential to discover ``unknown unknowns'', objects we have not yet imagined or anticipated. The advance in sensitivity, wavelength coverage, and spectral resolution (in comparison to Spitzer imaging) is so great that we are almost guaranteed to run into surprises. Covering ultra-deep parameter space with imaging and spectroscopy early in \JWST's mission helps to ensure ample time for follow up studies in future cycles.

\begin{figure}[t]
    \centering
    \includegraphics[width=0.49\textwidth]{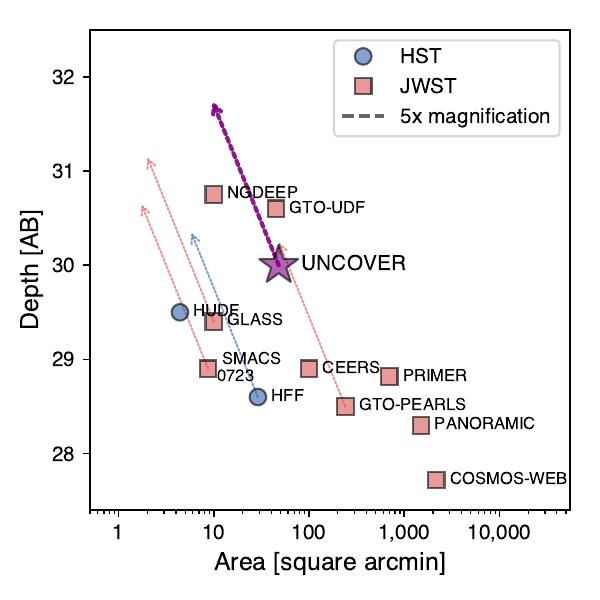}
    \caption{The UNCOVER combined imaging (purple star) probes a unique regime in the context of \HST extragalactic ultradeep fields (blue circles) and \JWST Cycle 1 imaging surveys (red squares); without lensing it probes deeper than previous \HST surveys and wide field programs. Gravitational lensing (approximate lensing vectors indicated by dashed arrows) from the Abell 2744 allows UNCOVER imaging to probe the intrinsically faintest objects of any \JWST project in Cycle 1. We note that lensing vectors are approximate, as the UNCOVER survey includes lensed and unlensed areas of Abell 2744.}
    \label{fig:deptharea}
\end{figure}

In Figure \ref{fig:deptharea} we show the full area of the UNCOVER survey (purple star) in context of other \HST (blue circles) and \JWST Cycle 1 extragalactic deep fields (red squares). UNCOVER is more than a magnitude deeper than other wide-field surveys. Only the ultradeep NGDEEP \citep[PIs: Finkelstein, Papovich, and Pirzkal, JWST-GO-2079,][]{bagley:24} survey and GTO ultradeep field  {(JADES, PI:Eisenstein, JWST-GTO-1180)} probe beyond UNCOVER, and these depths do not account for the effects of gravitational lensing. The Abell 2744 lensing cluster boosts the intrinsic limits of UNCOVER, sacrificing area (see arrows indicating the effect of a factor of ${\sim}5$ magnification in cluster fields) to make it the deepest extragalactic field in Cycle 1 of \JWST observations. Furthermore, UNCOVER is the only GO dataset that  {collected} deep spectroscopic followup of \JWST-selected objects with no proprietary time in Cycle 1.

The outline of this paper is as follows. The survey design of the UNCOVER program is presented in \S 2. This paper includes the public release of early reduced mosaics of NIRCam cluster pre-imaging and NIRISS parallel imaging, described in \S3. We summarize the science cases that will be enabled by the UNCOVER dataset in \S4. Finally, we summarize the prospects of the UNCOVER survey in \S5. Throughout this paper we adopt a standard $\Lambda$CDM cosmology with $\mathrm{H_0=70\,km\,s^{-1}Mpc^{-1}}$, $\Omega_M=0.3$, and $\Omega_{\Lambda}=0.7$.

\begin{figure*}[t]
    \centering
    \includegraphics[width=\textwidth]{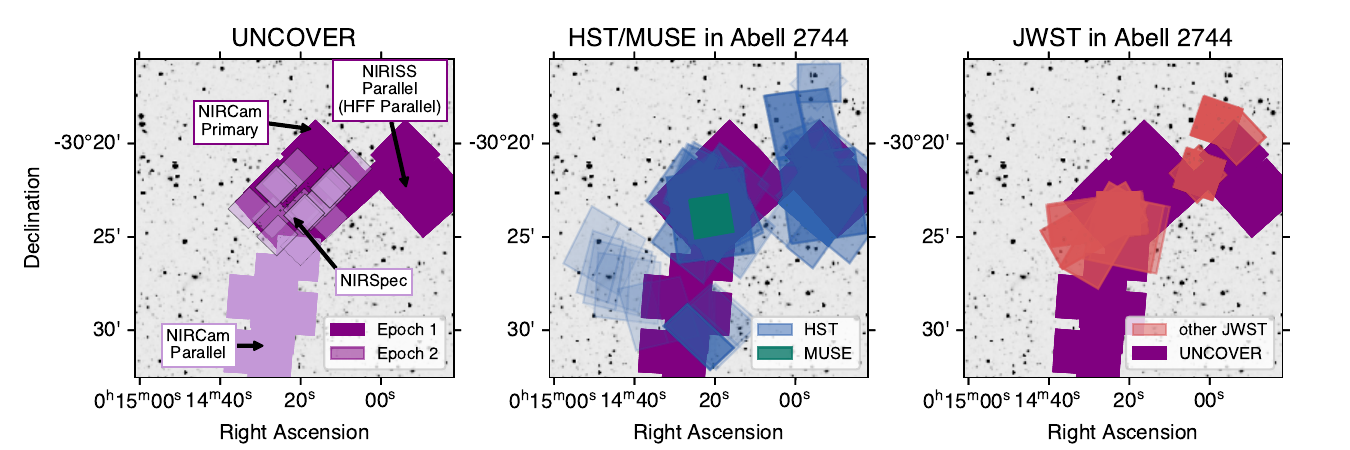}
    \caption{The Abell 2744 cluster has extensive deep optical/NIR (\HST and \emph{VLT}/MUSE) and \JWST coverage. The \JWST/UNCOVER footprints (purple) along with \HST imaging (blue) and \emph{VLT}/MUSE deep datacube (green, center) and \JWST Cycle 1 imaging and spectroscopy (red, right) against imaging from the Legacy Imaging survey \citep{dey:19}.  {An updated version with Cycle 2 JWST programs can be found in \citet{suess:24}.} In the left panel, we highlight the detailed layout of the UNCOVER dataset, differentiating between the first epoch (dark purple, NIRCam primary imaging and NIRISS parallel imaging) and second epoch (planned NIRSpec spectroscopy and NIRCam parallel imaging, light purple). The NIRCam mosaic is designed to match the $\mu=2$ magnification curve (see Fig.\ref{fig:magnification}). The footprints of the NIRSpec spectroscopy are provisional and subject to target selection.}
    \label{fig:footprints} 
\end{figure*}

\section{Survey Design}

\subsection{Observational Strategy}
The first UNCOVER data consist of deep (4{-}6 hour per filter) NIRCam pre-imaging on the A2744 cluster to ${\sim}29.5{-}30$AB magnitude in 7 filters, collected in November 2022. Nine months later  {July/August 2023} we  {targeted} sources detected in NIRCam with ultra-deep 19 hour NIRSpec/PRISM low-resolution spectroscopy.  {Due to technical issues, a fraction of NIRSpec observations will be repeated in July/August 2024)}. Both sets of observations include deep parallel imaging (in NIRISS and NIRCam, respectively), to increase the area for deep photometric studies of high-redshift galaxies at mild ($1.1{-}1.3\times$) lensing magnification. The footprints of these components are shown in Figure \ref{fig:footprints} (left panel), and reproduced along with ancillary data from \HST and \emph{VLT}/MUSE (center panel), as well as other Cycle 1 \JWST data (right panel). 

\subsection{Field selection} 

\begin{figure}[t]
    \centering \includegraphics[width=0.49\textwidth]{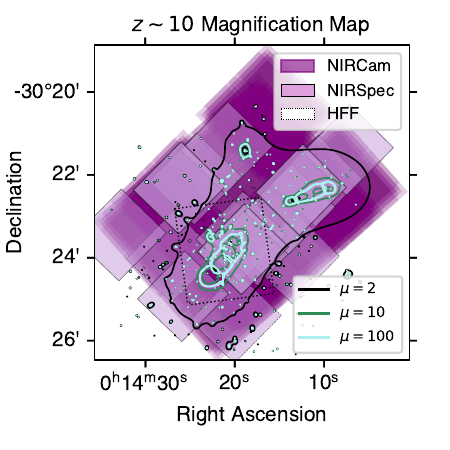}
    \caption{Gravitational lensing magnification contours in Abell 2744 are extremely extended, due to the complex structure of the multiple cluster cores.  {The above curves are taken from the UNCOVER-based lensing model at $z\sim10$ \citep{furtak:23lensing}}. The UNCOVER NIRCam mosaic (dark purple) spans the $\mu=2$ curve, which is significantly larger than the Hubble Frontier Field (black dashed outline). By extending to the northern subclumps the UNCOVER mosaic {enables} a more detailed mapping of that region. }
        \label{fig:magnification}
\end{figure}

A2744 is one of the most powerful known gravitational lensing clusters, with a large area of high magnification ($\mu{>}2,4,10$ over ${>}17,7,2$ arcmin$^2$), because the cluster itself consists of several merging subclusters, or massive subclumps \citep[e.g.,][]{merten:11,richard:14,arXiv:1504.02405,jauzac:15,diego:16,kawamata:17}. The full complex stretches towards the North-West of the Hubble Frontier Field cluster pointing. The NIRCam footprint is designed to span the $\mu>2$ magnification contour (see Figure \ref{fig:magnification}). The cluster contains many HST-detected $6{<}z{<}10$ galaxies, underscoring its efficacy. The field has excellent roll angle visibility for \JWST and low infrared background (ideal for deep background-limited observations). The observing windows (November 2022 for the imaging and expected July 2023 for spectroscopy) benefit from the lowest possible backgrounds in Cycle 1. The field also contains some of the best complementary multi-wavelength data (including deep $29$AB \HST/ACS optical data) and is accessible by the Atacama Large Millimeter/submillimeter Array (ALMA); we summarize ancillary datasets in \S \ref{sec:ancillary}. 

\subsection{Deep NIRCam and NIRISS imaging}

The UNCOVER survey imaging includes deep primary NIRCam imaging on the extended cluster and NIRISS and NIRCam parallel imaging in the outskirts. In this section, we describe the NIRCam primary field cluster pre-imaging mosaic and the NIRISS parallel. 

The NIRCam mosaic is designed to maximize the number of detected $z{>}10$ galaxies. This was done by forward-modeling the theoretical luminosity functions of \citet{mason:15} and from the \texttt{DELPHI} model \citep{dayal:14, dayal:22}  using the CATS v4.1 lens model of A2744 \citep{jauzac:15} to predict the number of $6{<}z{<}16$ galaxies to our detection limit. The number of $z{>}10$ galaxies is maximized by a 4-pointing gap-filled NIRCam mosaic. Our expected $5\sigma$ depths of ${\sim}$30 AB (given ${\sim}4$ hours of exposure in F200W and ${\sim}6$ hours in F115W, \JWST ETC 2.0, see Table \ref{tbl:exposures}) correspond to $M_{UV}{=}\minus 14.0$ at $z{=}6{-}7$ with ${<}3$ magnitudes of lensing \citep[where models are considered more robust, e.g.,][]{livermore:17,Atek:18,bouwens:21}. The dominant uncertainty in expected numbers is the difference between theoretical models, which can differ by factors of $>10$ at $z>10$ \citep[e.g.][]{Oesch:18,finkelstein:22b}, with a smaller contribution from cosmic variance ranging from $10\%-40\%$ at $z=7-10$ \citep[e.g.][]{Ucci:2021}. These depths compare favorably to those achieved in our final mosaic (see Table \ref{tbl:exposures}).

\begin{figure*}[t]
    \centering
    \includegraphics[width=\textwidth]{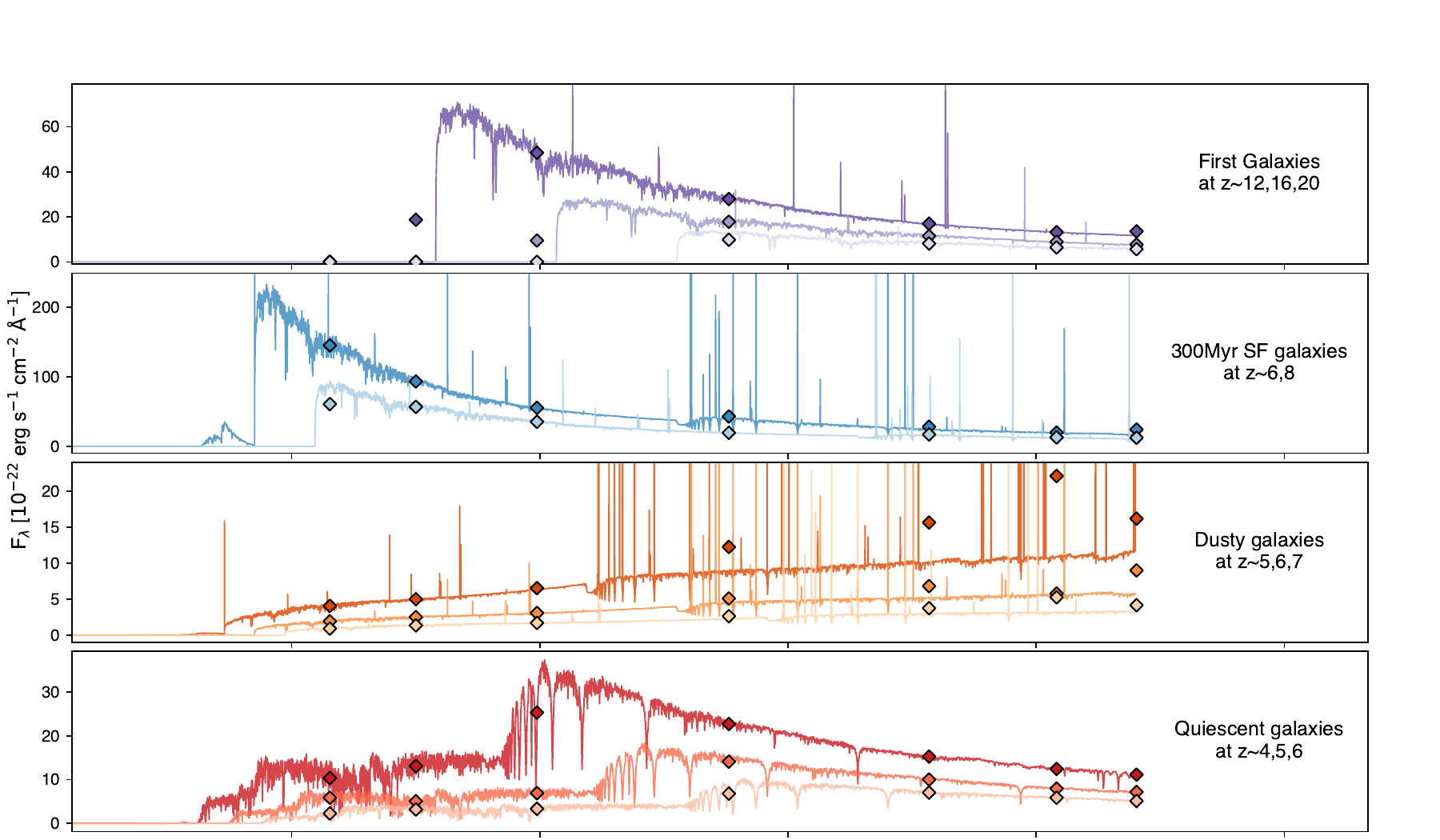}
    \includegraphics[width=\textwidth]{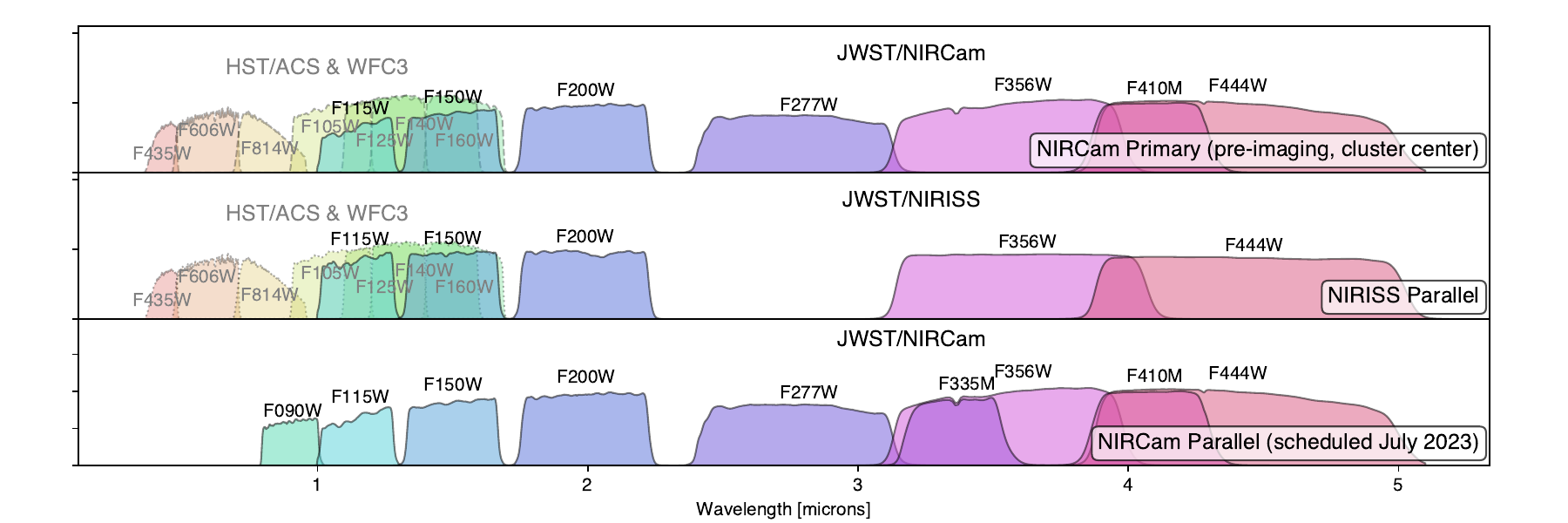}
    \caption{Top Rows: model SEDs for four key galaxy types and Bottom Rows: Filter curves for deep \HST and \JWST imaging in the 3 UNCOVER imaging fields. \HST imaging (ACS and WFC3) from the Hubble Frontier Field (HFF) program overlaps with the Abell 2744 cluster center (UNCOVER NIRCam primary, top row) and the HFF parallel imaging overlaps with the UNCOVER NIRISS parallel (middle panel). The NIRCam parallel lacks deep optical imaging from \HST, but includes F090W and \emph{two} medium band filters (F335M and F410M).
    }
    \label{fig:filters}
\end{figure*}

NIRCam pre-imaging is taken with a 4-point mosaic and an 8 pointing INTRAMODULEX dither pattern for 3.7-6 hours, using six broadband filters (F115W, F150W, F200W, F277W, F356W, and F444W). A medium band filter F410M is added, to improve diagnostics of emission lines and improve photometric redshifts and stellar masses of high-z galaxies \citep{Kauffmann:20,roberts-borsani:21,labbe:22}.  {We note that the same extended mosaic was followed up with JWST/NIRCam imaging as part of programs JWST-GO-4111 (PI Suess) and JWST-GO-3516 (PIs Matthee and Naidu), which added bluer broadband filters (F070W and F090W) and all remaining medium band filters.} Exposure times and approximate imaging depths are listed in Table \ref{tbl:exposures} and filter curves are shown in Figure \ref{fig:filters}. In that Figure, we show representative SEDs generated using \texttt{Bagpipes} \citep{carnall:18} with the NIRCam primary filters. We show delayed tau models ($\log M_{\star}/M_{\odot}=9.2$) in the top four panels (100 Myr-old in purple for the earliest galaxies, and 300 Myr-old in blue for the $z\sim6-8$, $A_V=3$ dusty galaxies in orange, and 500-Myr-old quiescent galaxies in red). NIRISS parallel imaging includes only F115W, F150W, F200W, F356W, and F444W broad bands. This field lacks some filters with respect to the NIRCam primary imaging as NIRISS lacks NIRCam's dichroic. The NIRISS parallel imaging fortuitously overlaps with 42-orbit $29 \mathrm{AB}$ F814W Hubble Frontier Field A2744 ACS parallel field, obviating the need for additional optical data. The first NIRCam/
NIRISS imaging data were collected on November 2, 4, 7, and 15, 2022. The final visit was a repeat observation of visit 1:1, which initially failed guide star acquisition (on October 31, 2022). The new visit 3:1 was observed at a slightly different angle ($\mathrm{V3PA}=45.00$ degrees relative to $\mathrm{V3PA}=41.3588$ degrees for the other 3 visits) and higher background level to facilitate rapid rescheduling. This results in a slightly askew mosaic footprint, but image depths and footprint are minimally impacted by this change.

In parallel to the follow-up NIRSpec observations, a second imaging parallel field will be taken with NIRCam using a more expanded set of seven broadband filters (F090W, F115W, F150W, F200W, F277W, F356W, F444W) and two medium band filters (F335M and F410M) with total exposure times of 2.6-5.3 hours. The expanded filter set mitigates the lack of optical data in that pointing. The NIRCam parallel observations are taken with the NIRSpec/PRISM in MOS mode with a 3-POINT-WITH-NIRCam-SIZE2 dither pattern  {and a three slitlet nod pattern}. 

\setlength{\tabcolsep}{2pt}
\begin{deluxetable}{ccccccc}[t]
\tabletypesize{\footnotesize}
\tablecolumns{8}
\tablewidth{0pt}
\tablecaption{UNCOVER Imaging\label{tbl:exposures}}
\tablehead{
\colhead{Instrument} & \colhead{Filter} & \colhead{Exposure} & \colhead{Dates} & \colhead{5$\sigma$ depth} & \colhead{ETC}}
\startdata
NIRCam & F115W & 6.0h & Nov.\ 2,4,7,15, 2022 & 30.05 & 29.9 \\ 
(Primary) & F150W & 6.0h & & 30.18 & 30.1 \\
& F200W & 3.7h & & 30.12 & 30.0 \\
& F277W & 3.7h & & 29.75 & 29.5 \\
& F356W & 3.7h & & 29.79 & 29.5 \\
& F410M & 3.7h & & 29.03 & 28.8 \\
& F444W & 4.6h & & 29.25 & 29.2 \\
\hline
NIRISS & F115W & 3.7h & (see primary) & 30.19 & 30.1\\
(Parallel) & F150W & 3.7h & & 30.13 & 30.2\\
& F200W & 3.7h & & 30.25 & 30.2 \\
& F356W & 2.1h & & 29.40 & 29.2 \\
& F444W & 2.1h & & 28.8 & 28.6 \\
\hline
NIRCam & F090W & 5.3h &  July {/August} 2023 & --- & 29.7 \\
(Parallel) & F115W & 5.3h & & --- & 29.8 \\
& F150W & 5.3h & & --- & 30.0 \\
& F200W & 2.6h & & --- & 29.8 \\
& F277W & 2.6h & & --- & 29.3 \\
& F335M & 2.6h & & --- & 29.0 \\
& F356W & 2.6h & & --- & 29.3 \\
& F410M & 2.6h & & --- & 28.6 \\
& F444W & 5.3h & & --- & 29.2\\
\enddata
\tablecomments{Imaging depths in the NIRISS/NIRCam mosaics are calculated using 0.08" radius apertures in the short wavelength bands, 0.16" radius apertures in the long wavelength bands, based on the noise properties inferred from the weight maps and corrected to total assuming point sources. The depths correspond the two-visit depth regions of the mosaic. ETC
values correspond to S/N=5 point source depths using JWST ETC v2.0.  {We refer the reader to \citet{Weaver:23} for a more detailed discussion of effective depths of our photometric catalogs and \citet{price:24} in preparation for the NIRCam parallel imaging.}}
\end{deluxetable}
\setlength{\tabcolsep}{2pt}

\subsection{Scheduled NIRSpec/PRISM spectroscopic followup}

The spectroscopic component of the UNCOVER survey {were completed in July-August 2023, aside from a failed visit that was impacted by an electrical short. We describe the final spectroscopic program design and observational strategy in \citet{price:24}.} These spectra  {are designed to be} ultradeep, using the low-resolution ($R\sim 30-300$) PRISM mode to provide the deepest continuum depths and widest wavelength coverage (0.6-5.3 microns). A primary goal of UNCOVER is to detect continuum flux and to measure continuum redshifts of any faint high redshift object detected securely with NIRCam ($10\sigma, {\sim}29$AB). This requires SNR${=}3$ per resolution element at 1.5\micron, which can be reached in ${\sim}20$ hours integration. For galaxies with strong emission lines, we can measure redshifts and emission line strengths to ${\sim}30$AB if EW$_{\mathrm{obs}}{>}600\mathrm{\AA}\,$ corresponding to e.g., rest-frame EW$_0$(H$\alpha$,0)${>}100\mathrm{\AA}$ at $z{=}6$ or EW$_0$([OIII]$_{5007}){>}60\mathrm{\AA}$ at $z{=}9$. Typical galaxies at these redshifts have ${>}5\times$ stronger emission lines \citep[e.g.,][]{stark:12,labbe:13,smit:14,stefanon:21, stefanon:22,williams:22}; we expect to measure redshifts for the vast majority of targets found by NIRCam.

\begin{figure*}[t]
    \centering
    \includegraphics[width=0.95\textwidth]{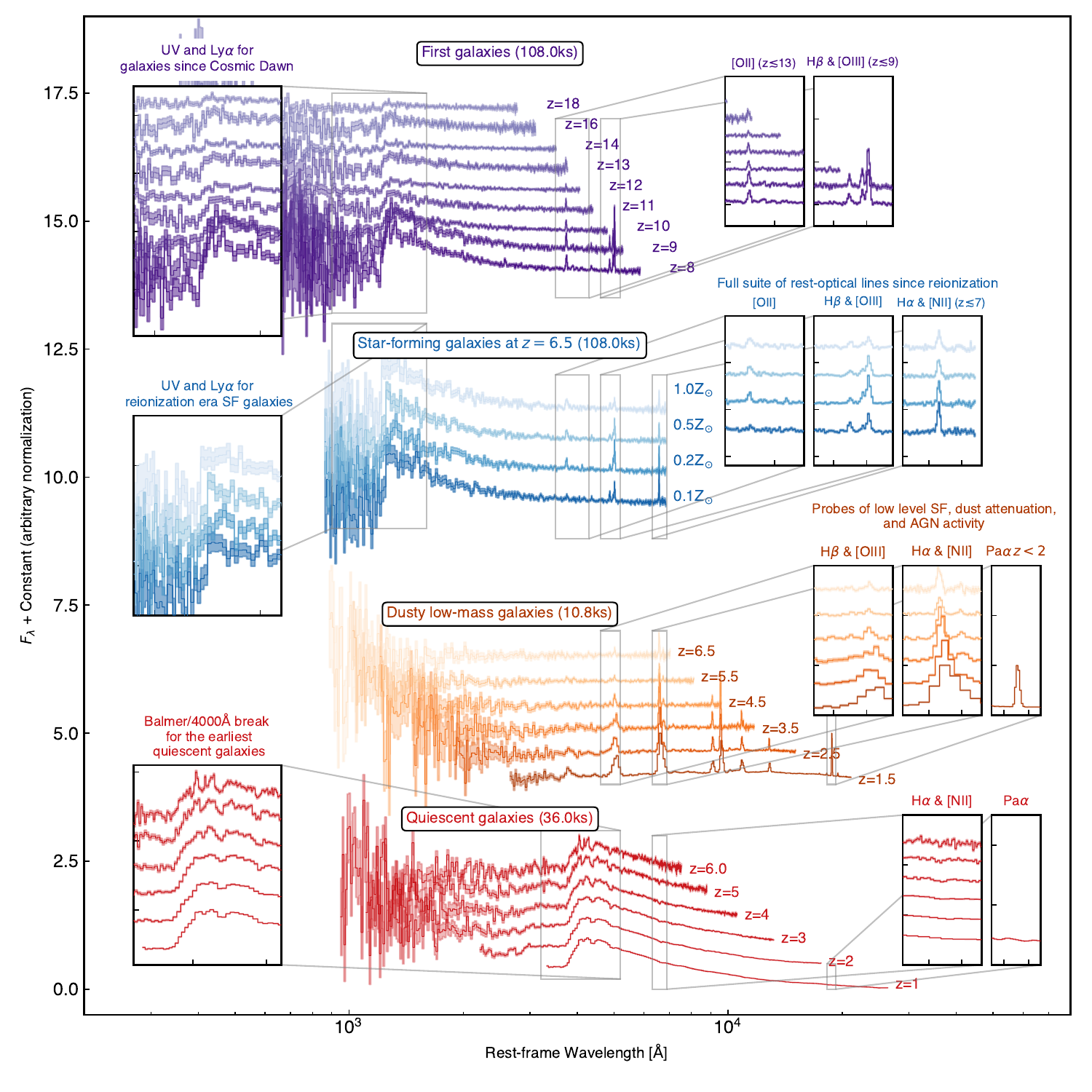}
    \caption{Simulated PRISM spectra of a variety of UNCOVER targets {, with similar SEDs to those presented in Figure \ref{fig:filters}}, with total exposure times ranging from 2.7-17.4 hours. The wide wavelength coverage (0.6-5.3 microns) of the NIRSpec/PRISM spectra catch critical spectroscopic features, with a resolution up to $R\sim300$ at the red end. Ultradeep exposures  {probe} the continuum flux for nearly all sources and the multiple mask designs  {provide} spectra for $\sim500-1000$ targets.}
    \label{fig:spectra}
\end{figure*}

The broad wavelength coverage  {spans} critical spectral features for all potential targets. Simulated spectra for a number of targets are shown in Figure \ref{fig:spectra}. From Cosmic Dawn through reionization, our PRISM spectra will probe UV flux and Ly$\alpha$ and will include rest-frame optical emission lines (including H$\alpha$ and [NII] below $z\sim7$). Deep exposure times probe rest-optical emission lines to measure low-level star-formation, dust attenuation, and AGN activity in dusty galaxies. Finally, the Balmer absorption features and Balmer/4000\AA{} breaks  {can} be clearly detected for even the most distant quiescent candidates.  {We include a similar figure with real spectra in \citet{price:24}.}

\begin{deluxetable}{ccc}[t]
\tabletypesize{\footnotesize}
\tablecolumns{8}
\tablewidth{0pt}
\tablecaption{Predicted Galaxy Counts in UNCOVER}
\tablehead{
\colhead{Category} & \colhead{Lensed (cluster)} & \colhead{Parallel}
}
\startdata
$z\gtrsim 12$ candidates & 36 & 9 \\
$9<z<12$ candidates & ${\sim}100$s & ${\sim}50$ \\
$6\lesssim z \lesssim 7$ galaxies & ${\sim}1500$ & ${\sim} 1500$\\
$z>1$ Dusty star-forming galaxies & ${\sim}50$ & ${\sim} 150$\\
$z>3$ Quiescent galaxies & ${>}6$ & ${>}10$ \\
$1<z<3$ Quiescent galaxies & ${>}20$ & ${>}60$ 
\enddata 
\tablecomments{The photometric dataset inevitably  {yielded} other extraordinary targets that  {have been} included in the MSA designs including, but not limited to: black hole seeds, extremely lensed galaxies and stars. All numbers are rough estimates derived from the JAdes extraGalactic Ultradeep Artificial Realizations (JAGUAR) Mock Catalog \citep{williams:18} below $z<6$ and at higher redshifts from the \citet{mason:15} model. Spectroscopic followup  {prioritizes} highly lensed or otherwise remarkable objects.}
\label{tbl:numbers}
\end{deluxetable}

Our planned spectroscopic targets will be roughly prioritized according to scientific value and rarity:
 {
\begin{itemize}
\item any $z{>}12$ candidates
\item $z{>}9$ galaxies prioritized by brightness
\item Pop III candidate sources
\item faint highly magnified $6{<}z{<}7$ galaxies
\item quiescent galaxies $z{>}4$
\item $z{>}6$ AGN
\item dusty galaxies $z{>}4$
\item low mass quiescent galaxies at $1{<}z{<}6$
\item any unusual or unexpected sources
\item extreme emission line galaxies
\item mass-selected galaxies sampled in bins of mass and redshift
\end{itemize}
}
We estimate that we can accommodate $\sim$15-20 sources to our full depth of 17.4 hours. Other sources require less exposure time.  {This is accomplished by switching individual sources in and out of overlapping MSA designs (described below).} Approximate numbers for key target classes are included in Table \ref{tbl:numbers}, estimated from \citet{williams:18} and \citet{mason:15} models.  {The full implementation of this MSA design will be presented in \citet{price:24}.}

 {The total NIRSpec integration times are designed to be split up in 7 partially overlapping dithered sequences of 2.7-4 hours each. We therefore design 7 masks with exposure time ranging from 2.7-17.4 hours, repeating the high priority objects. Each of our seven micro-shutter array (MSA) configurations are designed in an iterative manner, with the possibility of keeping objects on multiple masks. } Observations will follow a 3 point dither {pattern} (3-POINT-WITH-NIRCAM-SIZE2). We adopt the NRSIRS2 readout pattern, averaging 5 frames into groups to optimize the noise characteristics.
Given the high target density of some lower priority targets (e.g., there will be 1000s of high-z emission line galaxies), we expect to fill each mask with $\sim$100 targets for a total of $\sim$500 unique sources in the spectroscopic sample.

Slit loss through the MSA will be significant and wavelength dependent due to variation in the PSF, complicating precise flux calibration. However, because the UNCOVER program includes multiband NIRCam pre-imaging, our team  {is} well-positioned to perform corrections for these wavelength-dependent aperture effects, leveraging flexible, spatially-resolved galaxy models \citep{leja:21}.

\subsection{Ancillary Datasets in Abell 2744}\label{sec:ancillary}

\setlength{\tabcolsep}{0pt}
\begin{deluxetable*}{cccc}[t]
\tabletypesize{\footnotesize}
\tablecaption{Abell 2744 Ancillary Data\label{tbl:ancillary}}
\tablehead{\colhead{Target} & \colhead{Instrument} & \colhead{(Survey/)Program ID/PI} & \colhead{Notes/Depth}}
\startdata
\\[-5pt]
\multicolumn{4}{c}{\bf\JWST  {Cycle 1}}\\
\hline\hline\\[-5pt]
\multirow{6}{*}{Cluster (HFF-core)} & NIRSpec & \multirow{2}{*}{GLASS/ERS-X1324/PI: Treu} & 52ks \\
                      & NIRISS & & 35ks\\[+5pt]
                      & NIRCam & \multirow{2}{*}{DD-2756/PI: Chen} & \multirow{2}{*}{17ks}\\
                      & NIRSpec & & \\[+5pt]
                      & NIRCam & GTO-PEARLS/GTO-1176/PI: Windhorst & 12ks \\[+5pt]
\hline\\[-5pt]
\multirow{3}{*}{Field} & {NIRCam} & \multirow{3}{*}{GLASS/ERS-Treu/PI: Treu} & \multirow{3}{*}{30ks,50ks}\\
                      & F090W, F115W, F150W, F200W \\
                      & F277W, F356W, F444W \\
\\[-5pt]
\hline\hline
\multicolumn{4}{c}{\bf \HST}\\
\hline\hline\\[-5pt]
\multirow{7}{*}{Cluster (HFF-core)} & ACS/F435W & \multirow{3}{*}{\#11689/PI:Dupke, \#13386/PI: Rodney} & 18 orbits\\
                      & ACS/F606W &                                                       & 9 orbits\\
                      & ACS/F814W &                                                       & 41 orbits\\[+5pt]
                      & WFC3/F105W & \multirow{4}{*}{\#13495/PI:Lotz}                     & 24.5 orbits\\
                      & WFC3/F125W &                                                      & 12 orbits\\
                      & WFC3/F140W &                                                      & 10 orbits\\
                      & WFC3/F160W &                                                      & 24.5 orbits\\[+5pt]
\hline\\[-5pt]
\multirow{3}{*}{Cluster (expanded)}   & ACS/F606W & \multirow{3}{*}{BUFFALO, \#15117, PI:Steinhardt}       & 2/3 orbit \\
                      & ACS/F814W &                                                       & 4/3 orbit \\
                      & WFC3/F105W,F125W,F160W &                                          & 2/3 orbit \\[+5pt]
\hline\\[-5pt]
\multirow{9}{*}{Field}   & ACS/F435W & \multirow{3}{*}{\#13386/PI: Rodney, \#13495/PI:Lotz, \#13389/PI:Siana} & 6 orbits\\
                      & ACS/F606W &                                                       & 4 orbits\\
                      & ACS/F814W &                                                       & 14 orbits\\[+5pt]
                      & WFC3/F105W & \multirow{4}{*}{\#13495/PI:Lotz}                     & 24 orbits\\
                      & WFC3/F125W &                                                      & 12 orbits\\
                      & WFC3/F140W &                                                      & 10 orbits\\
                      & WFC3/F160W &                                                      & 24 orbits\\[+5pt]
                      & ACS/F606W,F775W & \#17231, PI: Treu                               & 15 orbits \\[+5pt]
\hline\hline
\multicolumn{4}{c}{\bf\emph{VLT}}\\
\hline\hline\\[-5pt]
Cluster (HFF-core)        & MUSE     &   \citet{mahler:18}                                   & 3.5,4,4,5hr \\[+5pt]
\hline\hline
\multicolumn{4}{c}{\bf\emph{Subaru}}\\
\hline\hline\\[-5pt]
\multirow{2}{*}{Cluster (30'x30')} & {Suprime-Cam/$B,R_C$} & \multirow{2}{*}{\citet{medezinski:16}} & {27.58,26.83} \\
& $i',z'$  &   &   26.31,26.03AB\\[+5pt]
\hline\hline
\multicolumn{4}{c}{\bf \emph{X-ray}}\\
\hline\hline\\[-5pt]
\multirow{3}{*}{Cluster} &  \emph{XMM-Newton} & \#074385010, 385 PI: Kneib & 110ks \\
                       &  \emph{Suzaku} &  \#808008010,\citet{eckert:16} & 75ks \\
                       &  \emph{Chandra} & \#23700107, PI: Bogdan & 1028ks \\[+5pt]
\hline\hline
\multicolumn{4}{c}{\bf\emph{ALMA}}\\
\hline\hline\\[-5pt]
\multirow{2}{*}{Cluster}                  & 1.1mm imaging & ALMA\#2013.1.00999.S, PI: Bauer & 395.6s \\
                      & 15GHz spectral scan & ALMA\#2018.1.00035.L, PI: Kohno & 95.5h \\[+5pt]
                      Cluster (4'x6')        & 30GHz spectral scan & ALMA\#2022.1.00073.S, PI: Fujimoto & 37.2h \\[+5pt]
\enddata
\end{deluxetable*}
\setlength{\tabcolsep}{2pt}

Abell 2744 has been targeted by a multitude of \HST imaging, e.g., through the Hubble Frontier Fields program \citep{lotz:17} and the \HST/BUFFALO survey \citep{steinhardt:20}, which we enumerate along with a variety of other ancillary datasets in Table \ref{tbl:ancillary}. \HST/ACS imaging in the cluster center was taken by Program \#11689 (PI: Dupke) and \# 13386 (PI: Rodney) and \HST/WFC3 observations were collected through program \#13495 (PI: Lotz). In the Hubble Frontier Field parallel (UNCOVER NIRISS parallel), \HST/ACS imaging was taken by  Programs \# 13386 (PI: Rodney), \# 13495 (PI: Lotz), and \#13389 (PI: Siana) and \HST/WFC3 imaging in that field was taken again by the Hubble Frontier Field program (\#13495, PI: Lotz). Although these deep \HST observations were limited to the field of view of individual ACS or WFC3 pointings, the footprints were expanded by the BUFFALO survey \citep[Program \# 15117, PI: Steinhardt;][]{steinhardt:20}, which expands the main cluster and parallel footprints by a factor of four. Recently, the deep optical coverage was expanded by Program \#17231 (PI: Treu) {\citep{paris:23}}. Amongst other things, this wide area enables improved lens models.

In addition, the Abell 2744 cluster has been targeted by ground-based imaging and spectroscopic programs. One such rich dataset is provided by deep \emph{VLT}/Multi Unit Spectroscopic Explorer (MUSE) observations. This mosaic of MUSE pointings covers 2' x 2' and yields untargeted spectroscopic redshifts for 514 galaxies \citep{mahler:18}. These spectroscopic redshifts contribute to improved lensing models, especially for multiply imaged systems. \citet{mahler:18} suggest that this improvement is approximately a factor of $\sim2.5$. Although Spitzer imaging in Abell 2744 is largely superseded by novel \JWST imaging from the UNCOVER survey, the field has also been observed in the X-ray for 110 ks by \emph{XMM-Newton} (\#074385010, PI: Kneib), 75 ks with \emph{Suzaku} \citep{eckert:16}, and \emph{Chandra} (\#23700107, PI Bogdan). Deep ground-based optical imaging exist from \emph{Subaru/Suprime-Cam} in \emph{B, R$_C$, i'}, and \emph{z'} \citep{medezinski:16}. ALMA observations of the Abell 2744 cluster include 1.1mm imaging of the cluster through the HFF-ALMA program \citep{gonzalez-lopez:17} and a 15GHz-wide spectral scan at 1.2mm through the ALMA lensing cluster survey (ALCS) \citep{kohno:19}.

Already in the first cycle of \JWST observations,  {several} other programs  {targeted} Abell 2744. First, the Early Release Science (ERS) GLASS-JWST program \citep[][PI: Treu]{treu:22} targets the cluster center with deep targeted NIRSpec spectroscopy and NIRISS imaging and untargeted spectroscopy and obtains NIRCam parallel imaging in the cluster outskirts in the following broadband filters (F090W, F115W, F150W, F200W, F277W, F356W, F444W). For details about that program, the reader is referred to the survey design paper \citep{treu:22}. All data from the GLASS-\JWST are available with no proprietary period. Additionally, Abell 2744 will be observed with NIRCam imaging as part of the \JWST GTO PEARLS (Prime Extra-galactic Areas for Reionization and Lensing Science) program 1176 (PI: Windhorst), including $\sim2$ hour depth observations in F090W, F115W, F150W, F200W, F277W, F356W, F410M, and F444W to $\sim28-29$AB 5$-\sigma$ limiting depths \citep{windhorst:22}. Recently, a DDT program (\#2756, PI: Chen) was approved to image and spectroscopically image a lensed, $z\sim3.5$ supernova. That dataset includes two epochs of NIRCam imaging (F115W, F150W, F200W, F277W, F356W, and F444W) and NIRSpec/PRISM spectra in the cluster center.   {A number of subsequent JWST programs have been completed in Abell 2744; we refer the reader to \citet{suess:24} for an updated accounting and similar figures.} Finally 4' x 6' of the cluster center (in the footprint of the UNCOVER NIRCam imaging)  {has been} mapped by  {the Deep UNCOVER-ALMA
Legacy High-z (DUALZ) Survey} ALMA band 6 spectral scans  {\citep{fujimoto:23dualz}}. 

\begin{deluxetable*}{ccc}[t]
\tabletypesize{\footnotesize}
\tablecolumns{8}
\tablewidth{0pt}
\tablecaption{UNCOVER Data Release Schedule\label{tbl:DR}}
\tablehead{
\colhead{Data Release} & \colhead{Description} & \colhead{Anticipated Date}}
\startdata
\\[-5pt]
\multirow{8}{*}{DR1/DR2} & Imaging Mosaics & \multirow{2}{*}{Dec. 2022} \\
& (this paper)  \\
& Updated Gravitational Lens Model & \multirow{2}{*}{Dec. 2022} \\
&  {\citep{furtak:23lensing}}  \\
& First-look Photometric Catalogs & \multirow{2}{*}{Dec. 2022} \\
&  {\citep{Weaver:23}} \\
& Photo-$z$s and Stellar Populations & \multirow{2}{*}{Jan. 2023} \\
&  {\citep{wang:24catalog}}  \\[5pt]
\hline\\[-5pt]
\multirow{2}{*}{DR3} & Photometric Catalogs & \multirow{2}{*}{ {April 2024}}\\
&  {with MegaScience Medium Bands} \\
&  {\citep{suess:24}} &  \\[5pt]
\hline\\[-5pt] 
\multirow{2}{*}{DR4} & Reduced PRISM Spectra & \multirow{2}{*}{Aug. 2024} \\
&  {\citep{price:24}} \\[5pt]
\enddata
\end{deluxetable*}

\subsection{Planned Data Release Schedule}

While there is no proprietary period associated with the UNCOVER dataset, our team is committed to the regular public release of reduced and high-level data products to maximize community use of this Treasury program. This paper represents the first of several data releases of the UNCOVER survey (DR1); see full anticipated release schedule in Table \ref{tbl:DR}.  {We subsequently provided two incremental data releases (DR1 and DR2)} prior to the \JWST Cycle 2 proposal deadline, providing the community with an early photometric catalog based on archival \HST imaging and NIRCam pre-imaging in the Abell 2744 cluster  {\citep{Weaver:23}} in addition to a description and catalog of derived physical properties, including photometric redshifts and stellar population synthesis modeling with \texttt{Prospector} \citep{johnson:21}  {\citep{wang:24catalog}}.  Additionally, our  {team has published} lensing maps that have been updated with new JWST sources  {\citep{furtak:23lensing}}, magnification estimates and uncertainties from this model  {are} included in photometric catalogs.  {We introduced a third data release to incorporate the additional broad and medium band NIRCam imaging and NIRISS parallel fields along with all photometric data products \citep[DR3][]{suess:24}.}
Within a year of our second epoch of data acquisition, we plan to publicly release reduced NIRSpec/PRISM spectra  {\citep{price:24} and lensing maps that have been updated with new JWST spectroscopic redshifts  {DR4}}. 

\begin{figure*}[t]
    \centering
    \includegraphics[width=1.0\textwidth]{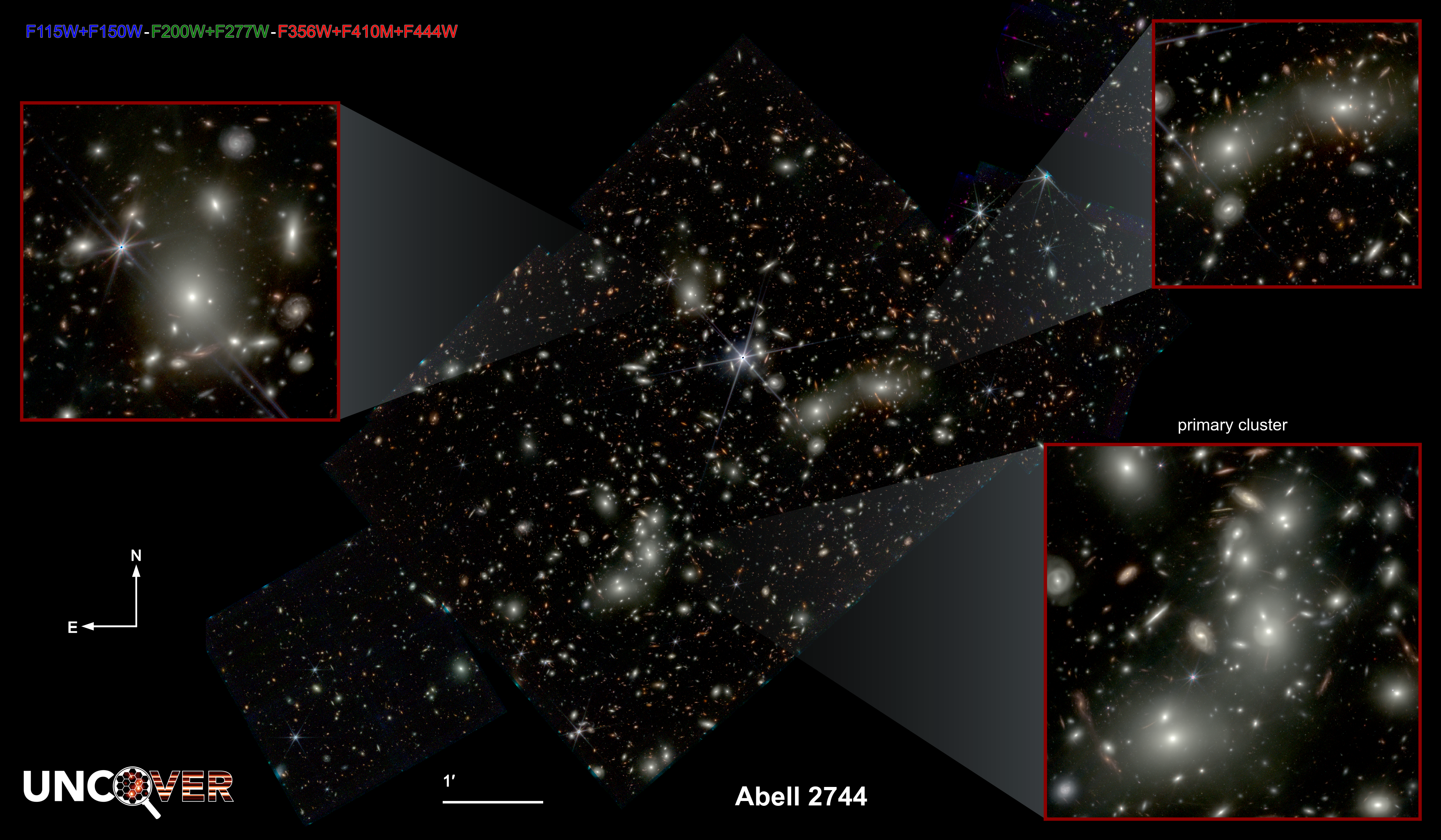}
    \caption{JWST imaging mosaic of Abell 2744, centered on the UNCOVER NIRCam mosaic. We highlight the three observed cluster cores in inset panels; only the Southern primary cluster has been covered by HST imaging in the field. Our imaging of the Northern substructures reveals a multitude of lensed features, that are used to improve the lens modeling in that region (L. Furtak et al., in preparation). This color image combines all UNCOVER filters and includes other JWST NIRCam/NIRISS imaging in the field including GLASS \citep{treu:22} and DDT\#2756. High resolution version of this color image is available on the UNCOVER website.}
    \label{fig:nircam}
\end{figure*}

\section{Image Reduction and Mosaics}

This paper focuses on the UNCOVER NIRCam and NIRISS imaging (on the Hubble Frontier Field parallel). A full description and public release of photometric and spectroscopic catalogs will be included in subsequent data releases and associated publications.

\subsection{Image Reduction}
All public \JWST NIRCam and NIRISS exposures in the Abell 2744 cluster  {field were downloaded from the Mikulski Archive for Space Telescopes  {The UNCOVER datasets can be accessed through MAST at \dataset[10.17909/zn4s-0243]{http://dx.doi.org/10.17909/zn4s-0243}, which includes data from the UNCOVER program as well as imaging from the GLASS-ERS (PI: Treu) and DD-2756 (PI: Chen) programs}}. The data reduction pipeline Grism redshift and line analysis software for space-based spectroscopy (\texttt{Grizli}, version 1.6.0.dev99) was used to process, align, and co-add the exposures. A detailed description of the pipeline is provided in G. Brammer et al.\ in preparation. 

 {We start with the MAST \texttt{rate.fits} products produced by Stage 1 of the \JWST calibration pipeline (v1.8.4) using calibration set \texttt{jwst\_0995.pmap}.  We derive a correction for the ``$1/f$ noise'' \citep{rauscher:15} by subtracting the source-masked median image values computed along detector rows and columns.  Bright cosmic rays that are not completely mitigated by the Level 1 exposure ramp fit can produce ``snowball'' residuals \citep{Rigby:22}.  Rather than re-running the ramp fits, we identify likely snowballs as large groups of pixels with the $\mathrm{DQ}=4$ bit set in the exposure data quality array and grow a conservative mask around them.}  

 {Before 2023 May, the flat-field reference files available in the JWST Calibration Reference Data System (CRDS) were determined from calibrations taken on the ground and did not accurately reflect the pixel-to-pixel structure in the in-flight NIRCam data.  This is most important in the reddest filters where the background is brightest and therefore where any errors in the multiplicative flat-field correction are largest.  We determined NIRCam ``sky flat'' reference files\footnote{\url{https://s3.amazonaws.com/grizli-v2/NircamSkyflats/flats.html}} from on-sky commissioning data from program COM-1063 (PI: Sunnquist) from normalized, source-masked exposures in each of the filter/detector combinations used in the UNCOVER field.  The NIRCam long-wavelength CRDS flat-field reference files were updated in 2023 May (\texttt{jwst\_1084.pmap}) based on in-flight data; however, we use the custom sky flats that were computed in a uniform way for all short- and long-wavelength filter/detector combinations.  We apply an additional mask to the exposure data quality arrays to ignore any pixels where the sky flat is outside of the range [0.7, 1.4], essentially removing approximately one out of four images of the same region on sky due to the dither pattern.  Any masked pixels will not contribute to the mosaic and the total effective exposure time at those positions will therefore decrease.  On average, 3.7\% of pixels are masked for one reason or another in the longest UNCOVER exposures (837~s in F277W), of which $\leq$~2.8\% come from the snowball masking.  After applying the flat-field calibration, we fit and subtract the additive ``wisp'' structure in some of the short-wavelength filter/detector combinations \citep{Rigby:22} using wisp templates\footnote{\url{https://s3.amazonaws.com/grizli-v2/NircamWisp/index.html}} derived from the GO-2561 UNCOVER data themselves. } 

 {The final step of the exposure-level processing is the photometric calibration, which was a rapidly-evolving topic of discussion at the time the UNCOVER data were taken in late 2022 \citep{boyer:22}.  This discussion was largely resolved with calibrations derived from in-flight data and provided by CRDS \texttt{jwst\_0989.pmap}.  However, we adopt a residual relative correction of 0.94 for the F277W filter in the NIRCam A module based on observations from the PRIMER program (GO-1837, PI: Dunlop) where the same sources were observed in the same filter on both detectors\footnote{See \url{https://github.com/gbrammer/grizli/pull/143}}. }

 {After the JWST-specific corrections to the exposures described above, the \texttt{grizli} pipeline processing is essentially identical to the procedure adopted earlier for data from the \textit{Hubble Space Telescope} \citep[e.g.,][]{kokorev:22}.  The most important step of this processing is the image alignment, which is performed in two steps.  The first computes shift translations based on the positions of sources identified in each individual exposure; these shifts tend to be small ($\lesssim0.1$ NIRCam pixel) as the pointing control and small offsetting of the telescope are quite precise \citep{Rigby:22}.  The absolute alignment is performed by aligning (shift and rotation) the UNCOVER F444W exposures to sources in the ground-based NOAO LegacySurvey DR9 catalog \citep{dey:19}.  We have verified that with this procedure the UNCOVER mosaic is aligned to the absolute frame defined by Gaia DR3 \citep{gaia:22} at a level of 12~mas \citep{Weaver:23}.    We then create a source list from the aligned F444W mosaic to which the exposures in all of the other filters are aligned.  The background pedestal level of each exposure are computed with the AstroDrizzle software \citep{Gonzaga:12}, which is then also used to create the final mosaics of each filter drizzled to a common pixel grid with drizzle parameters $\mathtt{pixfrac}=0.75$, $\mathtt{kernel}=\mathtt{square}$.\footnote{Note that \texttt{grizli} derives a world coordinate system (WCS) header from the JWST exposure metadata that is compatible with the Simple Imaging Polynomial convention \citep{shupe:05} so that the WCS from both JWST and HST exposures can be treated interchangeably.}  AstroDrizzle produces a ``weight'' image that has units of inverse variance of the ``science'' images and is made by propagating the total \textsc{err} noise model of each input exposure through the drizzle resampling.  The weight map does not account for correlated noise of the finite drizzle pixel resampling \citep[e.g., ][]{casertano:00}.  We also compute mosaics of all of the historical \textit{Hubble} imaging of the Abell 2744 field, drizzled to the same pixel grid and with datasets described by \cite{kokorev:22} and references therein.}

\subsection{DR1: Initial Release of NIRCam/NIRISS Mosaics}

\begin{figure*}[t]
    \centering
    \includegraphics[width=0.7\textwidth]{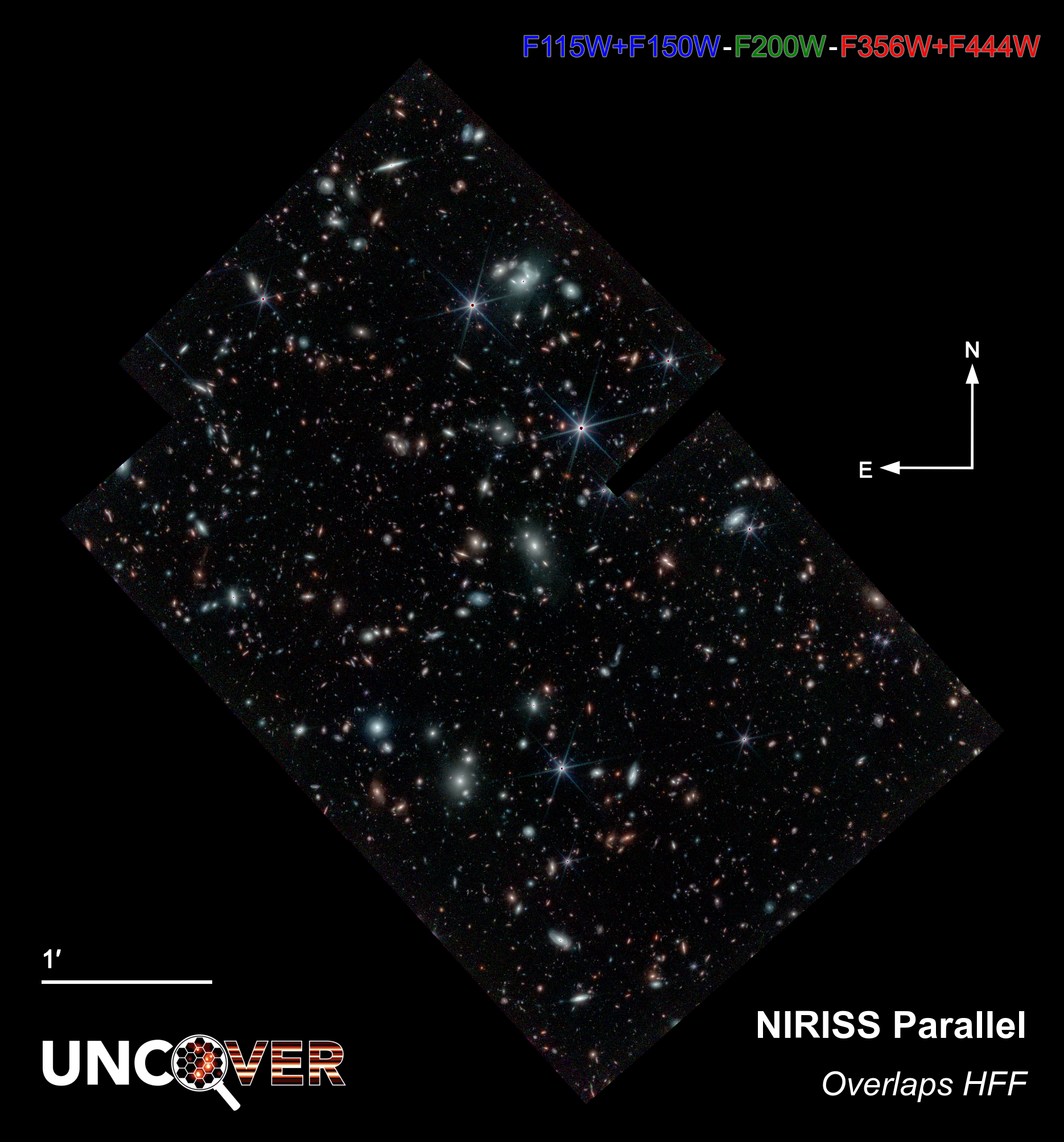}
    \caption{Color image of the UNCOVER NIRISS parallel, which overlaps with the Hubble Frontier field parallel.}
    \label{fig:niriss}
\end{figure*}

With this publication, we publicly release fully reduced, multiband mosaics for the first NIRCam cluster and NIRISS parallel imaging with 20 mas pixels in NIRCam short-wavelengths and 40 mas pixels in the NIRCam long-wavelength filters and NIRISS imaging. Color images for each mosaic are shown in Figure \ref{fig:nircam} (NIRCam) and Figure \ref{fig:niriss} (NIRISS). We also provide reduced \HST imaging on the same WCS grids. All of these data products will be hosted initially through Amazon Web Services, linked from 
the UNCOVER team website \footnote{\url{https://jwst-uncover.github.io/DR1.html}}, 
and upon publication uploaded to the Barbara A. Mikulski Archive for Space Telescopes (MAST). This data release is designed to be significantly in advance of the \JWST Cycle 2 proposal deadline. Details of this plan can be found in Table \ref{tbl:DR}.

Our team intends to incrementally release higher-level data products associated with these early data (e.g., intracluster light (ICL)-subtracted mosaics, photometric catalogs and derived properties, updated gravitational lensing maps) on a short timescale. These data releases will be followed by final catalogs and spectroscopic data releases in the subsequent years.

\section{Discussion and Scientific Objectives of the UNCOVER survey}
The \JWST UNCOVER survey is designed to address a primary objective of the observatory: detecting and spectroscopically characterizing the properties of the first galaxies, while enabling a broad range of scientific explorations. Broadly, our scientific targets fall into four primary categories.
\begin{figure*}[!t]
    \centering
    \includegraphics[width=\textwidth]{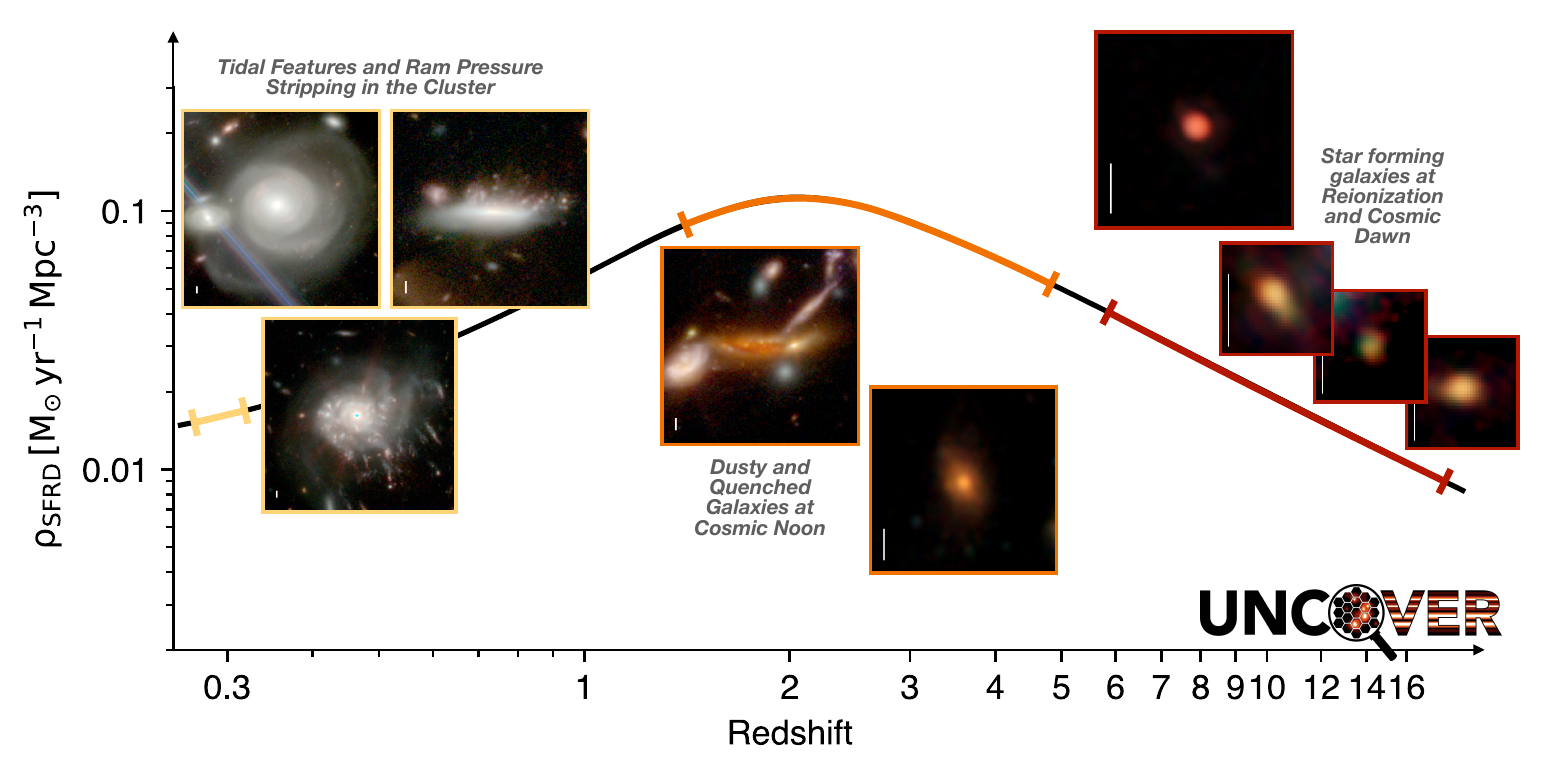}
    \caption{The UNCOVER imaging reveals spectacular views of a vast array of galaxies across cosmic time, shown here as drawn from the history of the cosmic star formation rate density from \citet{casey:18}. These galaxies span foreground  galaxies within the Abell 2744 cluster, to red and/or remarkably resolved lensed systems at Cosmic Noon (sub-mm galaxy A2744-ID02 from \citet{gonzalez-lopez:17} and a quiescent galaxy candidate), and back into the most extreme -- and therefore uncertain -- epochs to the time of the first galaxies (GLASSz8-1 from \citet{Roberts-Borsani:22} and triply-lensed JD1 from \citet{zitrin:14}). White scale bars indicate 0.5'' in all postage stamps.}
    \label{fig:gallery}
\end{figure*}

\begin{itemize}
    \item  {The First Galaxies:} The first \JWST observations released in July 2022  {quickly transformed} the field. Early Release Observations in SMACSJ0723.3-7327 \citep[ERO \#2736, PI: Pontoppidan,][]{pontoppidan:22} and Early Release Science observations from GLASS \citep[\#1324, PI:Treu,][]{treu:22} and CEERS \citep[\#1345, PI:Finkelstein][]{bagley:23} revealed an unexpected abundance of bright galaxies at $10 < z < 17$ \citep{naidu:22, donnan:22,atek:22,castellano:22, harikane:22, finkelstein:22, furtak:22} and a population of candidate massive ($>10^{10}M\odot$) galaxies at $7 < z < 10$ \citep{labbe:22}, which may or may not be in tension with galaxy formation model predictions \citep{mason:22,ferrara:22,boylan-Kolchin:22,nath:22}. The NIRCam imaging from UNCOVER extends these early efforts by providing deep 3.7 hour ($\sim$30 AB F200W) imaging over a large area with gravitational lensing ($\mu>$2 over $>17$ arcmin$^2$), ideally suited to improve statistics of galaxies at $z>10$ (see Table \ref{tbl:numbers}). The ultradeep $2.7-20$ hour $R{\sim}100$ PRISM spectra  {can} provide spectroscopic continuum redshifts of any \JWST-selected galaxy to 29AB, unbiased with respect to the presence of emission lines or targeting, providing the first comprehensive test of the photometric selection at these redshifts.  {The early UNCOVER spectra revealed a number of stunning examples of these incredibly distant galaxies \citep{wang:23distant,fujimoto:23firstgal}.}
    \item  {The Reionization Era:} Understanding how and when galaxies irradiated their environment and reionized the surrounding neutral gas is a key outstanding question.  To determine the contribution of galaxies to reionization requires  a full accounting of the faintest galaxies that dominate the integrated UV light, the rate at which they produce ionizing photons, and the fraction of photons that escape. UNCOVER imaging will securely detect galaxies to $M_{UV} {\sim} \minus 14$ and should settle the issue of the potential turnover  {\cite[e.g.,][]{2022ApJ...940...55B}}, directly measuring 85\% of the reionizing UV radiation  {\citep[e.g.,][and references therein]{livermore:17}}. The $3\times$ spatial resolution improvement of NIRCam versus \HST/WFC3 better constrains the size distribution of the galaxy population, further mitigating systematic sources of error. Ultra-deep PRISM spectra  {enables} for the study of ionizing spectra of the faintest galaxies  {(down to $M_{UV} {\sim} \minus 15$) \citep{atek:24}}. Of the expected ${\sim}$1500 galaxies at $z{=}6{-}7$, we estimate about ${\sim}40$ will have intrinsic UV magnitudes of $M_{UV}{>}\minus 16$ but still be bright enough (${<}28$AB) to get high quality spectra. For these galaxies we will observe the full complement of strong optical lines ([OII], [OIII], H$\beta$, H$\alpha$, [NII], [SII], with H$\alpha$ and [NII] marginally resolved at $R{\sim}300$ at ${\sim}5\micron$), to study ISM properties, ages, metallicities, and ionization mechanisms. 
    The combination of imaging and spectra has the potential to be a quantum leap in studies of reionization era galaxies.

    \item  {The Emergence of Dusty Galaxies:}   
    The fraction of star formation missed due to dust at $z{>}4$ and at low luminosities remains unclear.  ALMA has discovered dusty galaxies with relatively low masses and SFRs already at $z>5$ \citep[e.g.,][]{yamaguchi:19,wang:19,williams:19,fudamoto:21,algera:22, dayal:22}, suggesting that dust obscured star formation may be more prominent than was expected. These highly obscured extreme starbursts known to exist may represent only the tip of the iceberg. Detection of starlight from ``\HST-dark'' galaxies was expected in deep JWST imaging; early imaging has indeed revealed a  substantial population \citep{barrufet:22,nelson+22,perezgonzalez:22,price:23} of extreme red, dusty galaxies extending to $z\sim7-8$.  UNCOVER will place strong constraints at the lowest masses and the highest redshifts, complementary to the \JWST wide-field programs (e.g., COSMOS-Web \citep[GO \#1727, PI: Kartaltepe and Casey;][]{casey:22}, PANORAMIC (GO \#2514, PI: Williams and Oesch)) capable of finding massive obscured galaxies over larger volumes. The ability to detect any highly obscured $M_{\star}{>}10^{9}M_{\odot}$ galaxy to $z{=}9$ with NIRCam and test photometric stellar population inferences with NIRSpec spectroscopy will further expand our census of star formation. 
    
    \item  {The Epoch of Quenching:}
    Quenching of star formation is a complex phenomenon, with different processes dominating over a range of times, mass scales, and environments. The earliest massive quiescent galaxies have been spectroscopically confirmed to redshift $z{\sim}4$ \citep[e.g.,][]{glazebrook:17,schreiber:18,forrest:20}. However, this is entirely limited by detection: quiescent galaxies almost certainly exist to lower masses and higher redshift beyond the capabilities of \HST and ground-based spectroscopy. 
    The first \JWST imaging has already provided spectacular candidates of massive, quiescent galaxies to redshift $\sim5$ \citep{carnall:22}. UNCOVER NIRCam imaging will extend the search for quiescent galaxies to unprecedented low mass and high redshift, i.e., ${<}10^9M_{\sun}$ to $z{=}9$. \citet{marchesini:22} demonstrated the presence of $10^{10}M_{\sun}$ quiescent galaxies to $z\sim2.5$ with early NIRISS spectra. The UNCOVER NIRSpec/PRISM spectroscopy will be able to push this to even lower masses and higher redshifts due to the improved depth and wavelength coverage to $\sim5.3$ microns to confirm the quiescence through the absence of emission lines and strong Balmer breaks  {\citep[e.g.,][]{setton:24}}. Beyond identification and counting, the PRISM spectroscopy will also be deep enough to further constrain stellar populations to $F444W\lesssim 27.5$AB from the stellar continuum.
    
    \item  {The Unknown Unknowns:} Perhaps the most exciting legacy of deep fields like UNCOVER is the potential to discover objects that we have not yet imagined or identify predicted objects that we never hoped to detect in addition to offering possible constraints on the non-CDM nature of dark matter \citep[e.g.,][]{Dayal:15}. The advance in sensitivity, wavelength coverage, and spectral resolution is so large that we will certainly run into surprises. The UNCOVER treasury program ensures a comprehensive exploration of ultra-deep parameter space with public imaging and spectroscopy released early in \JWST's mission.  {An extraordinary example of this fruitful discovery space has been the abundance of actively accreting black holes that have been discovered and studied within the UNCOVER imaging \citep[e.g.,][]{furtak:23triple,greene:24lrd} and spectroscopy \citep[e.g.,][]{furtak:24triple,goulding:23,kokorev:23}.}
\end{itemize}

The astronomical community has only begun to scratch the surface of possible \JWST observations of the distant Universe. This paper aims to present a basic description of the UNCOVER Treasury program to optimize community use of early data products, prepare for Cycle 2 \JWST proposals targeting objects in and behind the Abell 2744 cluster, and anticipate use of the scheduled NIRSpec/PRISM follow-up. These studies are likely to range from detailed analysis of previously known sources to exciting \JWST discoveries. We conclude this paper by highlighting the broad range of UNCOVER scientific targets in Figure \ref{fig:gallery}. Individual objects are plotted against the backdrop of dust-rich and dust-poor models for the evolution of the star formation rate density from \citep{casey:18}.

Even in Cycle 1, the community  {had} access to a wide range of exciting \JWST datasets probing the distant Universe. Some of these programs will enable detailed high-resolution studies of extremely lensed sources, as in TEMPLATES (ERS \# 1355, PI: Rigby and Viera). The \JWST-GLASS program (ERS \# 1324, PI: Treu) provides an in-depth and multi-pronged view of reionization in the Abell 2744 cluster \citep{treu:22}. The CEERS program provides a range of relatively shallow imaging and spectroscopy across the well-studied CANDELS/EGS field \citep[ERS \#1324, PI: Finkelstein;][]{finkelstein:22b}. Other GO programs will expand imaging footprints: PRIMER (GO \#1837, PI: Dunlop) will map nearly 700 sq. arcmin with 10 NIRCam and MIRI filters (236 sq. arcmin in MIRI) and COSMOS-Web will map an incredible 0.6 square degrees in NIRCam and 0.2 square degrees in MIRI with a smaller number of filters \citep[GO \#1727, PI: Kartaltepe and Casey;][]{casey:22}. Finally, the pure parallel program PANORAMIC (GO \#2514, PI: Williams and Oesch) is planned to provide multi-band NIRCam imaging of spatially uncorrelated pointings covering up to 0.4 square degrees with a range of depths.

The UNCOVER program explores a unique parameter space; reaching within $\sim$ a magnitude of the GTO ultradeep field but boosted by gravitational lensing and with no proprietary access. Although this is just the first of several planned public data releases, our team looks forward to seeing what the community will uncover from this and all other \JWST treasure troves. 

\begin{acknowledgments}
\noindent
RB acknowledges support from the Research Corporation for Scientific Advancement (RCSA) Cottrell Scholar Award ID No: 27587. Cloud-based data processing and file storage for this work is provided by the AWS Cloud Credits for Research program. 
This work is based in part on observations made with the NASA/ESA/CSA \emph{James Webb Space Telescope}. The data were obtained from the Mikulski Archive for Space Telescopes at the Space Telescope Science Institute, which is operated by the Association of Universities for Research in Astronomy, Inc., under NASA contract NAS 5-03127 for \JWST. These observations are associated with \JWST Cycle 1 GO program{s} \#2561 {, JWST-ERS-1324, JWST-DD-2756}. Support for program JWST-GO-2561 was provided by NASA through a
grant from the Space Telescope Science Institute, which is operated by the Associations
of Universities for Research in Astronomy, Incorporated, under NASA contract NAS5-26555.  {This research is based on observations made with the NASA/ESA Hubble Space Telescope obtained from the Space Telescope Science Institute, which is operated by the Association of Universities for Research in Astronomy, Inc., under NASA contract NAS 5–26555. These observations are associated with programs HST-GO-11689, HST-GO-13386, HST-GO/DD-13495, HST-GO-13389, HST-GO-15117, and HST-GO/DD-17231.}  {The Cosmic Dawn Center is funded by the Danish National Research Foundation (DNRF) under grant \#140.}  {This research was supported in part by the University of Pittsburgh Center for Research Computing, RRID:SCR\_022735, through the resources provided. Specifically, this work used the H2P cluster, which is supported by NSF award number OAC-2117681.}
LF and AZ acknowledge support by Grant No. 2020750 from the United States-Israel Binational Science Foundation (BSF) and Grant No.\ 2109066 from the United States National Science Foundation (NSF), and by the Ministry of Science \& Technology, Israel. PD acknowledges support from the NWO grant 016.VIDI.189.162 (``ODIN") and from the European Commission's and University of Groningen's CO-FUND Rosalind Franklin program. HA acknowledges support from CNES (Centre National d'Etudes Spatiales). RS acknowledges an STFC Ernest Rutherford Fellowship (ST/S004831/1). MS acknowledges support from the CIDEGENT/2021/059 grant, from project PID2019-109592GB-I00/AEI/10.13039/501100011033 from the Spanish Ministerio de Ciencia e Innovaci\'on - Agencia Estatal de Investigaci\'on, and from Proyecto ASFAE/2022/025 del Ministerio de Ciencia y Innovación en el marco del Plan de Recuperación, Transformación y Resiliencia del Gobierno de Espa\~na. The work of CCW is supported by NOIRLab, which is managed by the Association of Universities for Research in Astronomy (AURA) under a cooperative agreement with the National Science Foundation.
\end{acknowledgments}

\vspace{5mm}
\facilities{\JWST{}(NIRCam, NIRSpec, and NIRISS), \HST{}(ACS and WFC3)}

\software{astropy \citep{2013A&A...558A..33A,2018AJ....156..123A,2022ApJ...935..167A}, Bagpipes \citep{carnall:18}, Grizli \citealp{brammer:2021}, Prospector \citep{johnson:21}, DrizzlePac \citep{Gonzaga:12}}

\end{document}